# Combined Microwave and Laser Rayleigh Scattering Diagnostics for Pin-to-Pin Nanosecond Discharges

**Xingxing Wang, Adam Patel, Alexey Shashurin**
Purdue University, 701 W Stadium Ave, West Lafayette, IN 47907, United States of America

**Abstract**
In this work, the temporal decay of electrons produced by an atmospheric pin-to-pin nanosecond discharge operating in the spark regime was measured via a combination of microwave Rayleigh scattering (MRS) and laser Rayleigh scattering (LRS). Due to the initial energy deposition of the nanosecond pulse, a variance in local gas density occurs on the timescale of electron decay. Thus, the assumption of a constant collisional frequency is no longer applicable when electron number data is extracted from the MRS measurements. To recalibrate the MRS measurements throughout the electron decay period, temporally-resolved LRS measurements of the local gas density were performed over the event duration. Local gas density was measured to be 30% of the ambient level during the later stages of electron decay and recovers at about 1 ms after the discharge. A shock front traveling approximately 500 m/s was additionally observed. Coupled with plasma volume calibration via temporally-resolved ICCD imaging, the corrected decay curves of the electron number and electron number density are presented with a measured peak electron number density of $4.5\times10^{15}$ cm$^{-3}$ and decay rate of ~ $0.1$-$0.35\times10^7$ s$^{-1}$. A hybrid MRS and LRS diagnostic technique can be applied for a broad spectrum of atmospheric-pressure microplasmas where a variation in number gas density is expected due to an energy deposition in the discharge.

Microwave Rayleigh scattering (MRS), or Rayleigh microwave scattering (RMS), has been proven as a powerful diagnostic for measuring total electron number (and number density) of microplasmas at atmospheric conditions - including helium plasma jets [1] [2] and laser induced plasmas [3]. The idea of MRS was first proposed by Shneider as a method for measuring total electron numbers in microplasmas [4]. The concept of the approach is based on the coherent scattering of microwave radiation by a small plasma volume in the quasi-Rayleigh regime. When the prolonged plasma object is oriented along the linearly-polarized microwave electric field, electrons in the filament are polarized and consequently emit radiation in a Hertzian-dipole fashion. The electric-field amplitude of the scattered radiation detected is thus proportional to the total number of electrons inside the plasma volume.

Studies on nanosecond discharges with a pin-to-pin geometry have been conducted utilizing the MRS technique [5] [6]. Nanosecond discharge plasmas have been employed in a variety of situations which require rapid gas heating or efficient production of metastable and active species. Namely, in applications ranging from boundary layer control [7] [8] to the improvement of flame stability [9] [10]. It is thus crucial to conduct detailed diagnostics of ns-discharge plasmas under various operating conditions. In previous MRS studies, the full temporal evolution of total electron number (and density) was measured for pin-to-pin discharges at different discharge conditions (gap distance, energy deposition, repetition frequency, etc.). In those studies, it was assumed that the signal from the MRS system was solely a function of the total electron number - where the local gas density, and consequently collisional frequency, can be considered constant throughout the entire decay process.

However, this assumption is only valid when 1.) the plasma decay process is fast so that the time frame of interest is too short for any gas density variation to be realized or 2.) the energy deposition from the discharge is sufficiently small such that any gas density disturbance is negligible. From our previous measurements, we observed that the overall total electron decay lasts for several microseconds and is thus susceptible to thermo/hydrodynamic influence (sound in air at standard atmospheric conditions travels about 0.3 mm within 1 µs which is comparable with a typical diameter of the discharge column [5]). In addition, the energy deposition per pulse is typically several millijoules, which results in a local gas temperature rise of several-thousand Kelvin [11]. As a result, the assumption of a constant collisional frequency must be validated and assessed through local bulk gas density measurements.

Laser Rayleigh scattering (LRS) is one potential means of determining collisional frequency through non-intrusive measurements of local gas density. The technique is well-established for structure, flow field, and gas density measurements in the study of aerodynamics, combustion, and plasma physics [12] [13] [14] [15]. In collective random-phase LRS, induced electric dipole radiation is scattered in an incoherent multi-body fashion in the non-forward direction – and is thus proportional to the gas density (number of scatterers). Scattering features additionally reflect perturbations in pressure, temperature, and internal energy states of the gas [12]. For atmospheric pressure plasmas where the ionization degree is relatively low, incoherent neutral Rayleigh scattering is expected to dominate over Thomson scattering on plasma electrons [16].

In this work, the MRS signal from a pin-to-pin discharge is recorded at atmospheric conditions. A laser Rayleigh scattering technique is then adopted to investigate local gas density variations throughout electron decay. The corrected total electron decay curve is correspondingly acquired utilizing a hybrid of the MRS and LRS techniques. With additional measurements of the plasma volume via an ICCD imaging technique, the temporal evolution of spatially-averaged electron number density is finally presented. The results of this study suggest the necessity of accurate local gas number density measurements for the correct correlation between MRS data and plasma electron number density.

A high voltage (HV) nanosecond pulse was applied to two tungsten pin electrodes separated by a gap distance $d = 5$ mm at atmospheric conditions ($T = 298$ K, $p = 1$ atm). The HV pulse was produced by an Eagle Harbor NSP-3300-20-F – with a peak value of 24 ± 0.1 kV and a pulse width of 90 ± 3 ns. The pulser was operated in single pulse mode at a maximum repetition frequency of 1 Hz to eliminate any memory effects between the adjacent pulses. Under these conditions, the discharge operates in the spark regime [6] [17] [18]. The corresponding average energy deposition per pulse is approximately 5 ± 0.2 mJ, calculated through numerical integration of the product between voltage and current. The voltage was measured by two high voltage probes (Tektronix P6015A), each connected to respective output legs of the pulser. The pulser was operated in a floated mode such that the voltage pulse applied to the electrodes was calculated as the difference between the two probe measurements. The discharge current was measured by a high-frequency current transformer (Bergoz FCT-028-0.5-WB). A schematic of the discharge setup and diagnostics is illustrated in *Figure 1* below and outlined in the color blue. Electric signals of various diagnostic equipment were interpreted through an oscilloscope (Lecroy HDO3904, 3 GHz bandwidth, 40 GS/s sampling rate).

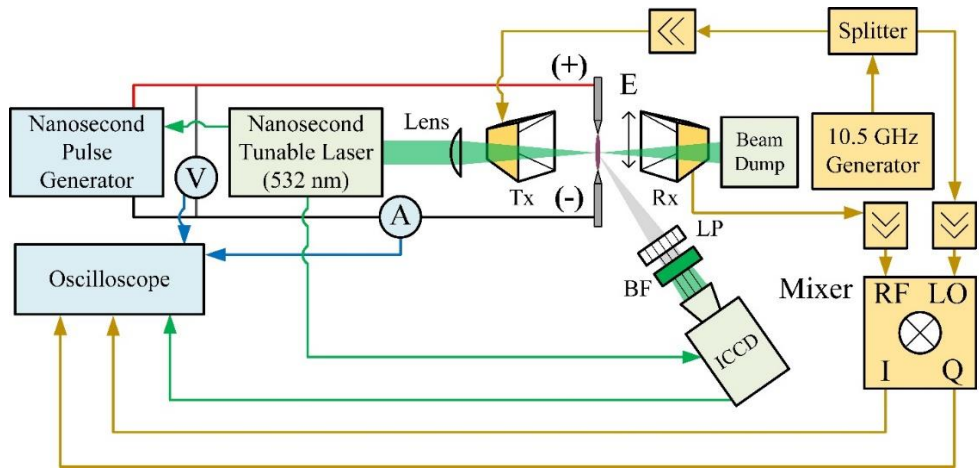

**Figure 1. Schematic of the ns-discharge setup (blue), MRS diagnostic equipment (gold), and LRS diagnostic equipment (green).**

The total electron number ($N_e$) and electron number density ($n_e$) is measured by the microwave Rayleigh scattering (MRS) technique and ICCD (Princeton Instruments PI-MAX 1024i) volume imaging. A 10.5 GHz MRS homodyne detection system was built and utilized for the absolute measurement of electron number density - after calibration with a dielectric material with known properties by Equation (1) below. The schematic of the MRS system circuitry is additionally depicted in *Figure 1* in yellow [4]. In this setup, the microwave horn antennas have a gain of 10 dB with a corresponding Tx output power of ~38 dBm.

$$U_{out} = \begin{cases} A\frac{e^2}{mv}N_e = A\frac{e^2}{mv}n_e V - & \text{for plasma} \\ AV\varepsilon_0(\varepsilon - 1)\omega & - \text{ for dielectric bullet scatterer} \end{cases} \quad (1)$$

where *e*: electron charge
  *m*: electron mass
  $\varepsilon_0$: dielectric permittivity of vacuum
  *ε*: relative permittivity of dielectric material
  *ω*: microwave frequency
  *v*: collisional frequency
  *A*: proportionality factor
  *V*: volume of the plasma/dielectric scatterer
  $U_{out}$: signal detected by the MRS system

The proportionality factor *A* is determined from the second expression in Equation (1) by applying the MRS system to a dielectric scatterer with known properties. It can then be applied to upper Equation 1 for the evaluation of the total electron number, $N_e$, in plasmas. The electron number density, $n_e$, can then be calculated using ICCD plasma volume measurements. It is important to note that there exists an upper limit on the plasma density that an MRS system can measure. In order for Equation 1 to be valid, the plasma diameter needs to be smaller than the skin layer depth - which is governed by the electron number density. For instance, for a skin layer depth of 300 μm, the soft upper limit of $n_e$ that a 10 GHz MRS system is applicable with is $10^{16}$ cm$^{-3}$.

 In addition to the aforementioned diagnostic techniques, a laser Rayleigh scattering system was utilized for the measurement of the local gas density, $n_g$, and determination of collisional frequency. As LRS is an incoherent scattering process, the intensity of the scattered signal is linearly proportional to the local gas density. It should be noted that the sensitivity of the laser Rayleigh scattering system is also affected by the cross section of the molecules, which is partially governed by the gas temperature. However, this dependence is relatively weak (~ 1% per 1000 K) [19]. Given that the temperature variation for the pin-to-pin ns-discharge is less than 5000 K, we thus assume that the change of the scattering signal is solely determined by the local gas density. A ~5 ns, 1 ± 0.1 mJ laser pulse at 532 ± 0.5 nm was produced by a broadly-tunable Nd:YAG laser (EKSPLA NT342) with Pellin-Broca spectral cleaning filter. The output laser pulse was found to be a linearly-polarized 7 mm dia. top hat profile – which was then focused to the center of the ns-discharge via a 175 mm plano-convex lens. The scattered signal was finally captured by the aforementioned ICCD camera, with an additional 532 ± 10 nm band pass filter, film polarizer, sufficiently short camera gate width, and broad spectra absorbing backdrop for LRS signal isolation (mitigate reflection, straylight, and fluorescence). 532 nm was selected as the central wavelength due to low absorption by neutral and ion species. It will be shown below that even though the LRS signal overlaps with Thomson and Raman scattering [16], the contribution of Rayleigh scattering is dominant for the experiemntal conditions in this work. In experiment, averaging of the scattering event at consistent timestamps was performed to mitigate variance in laser and ns-discharge features. Timing of LRS was achieved through external triggering of the nanosecond pulser and ICCD camera from the laser (laser jitter ± 0.5ns) and confirmed via a combination of the ICCD monitor and oscilloscope signals. A detailed schematic of the LRS system is additionally depicted in **Figure *1*** in green. Gas density $n_g$ is measured from 150 ns to 1 ms after the initiation of the HV pulse where the local density finally returns to ambient conditions.

A typical temporal evolution of the voltage (*V*) and current (*I*) waveforms are presented below in **Figure *2***. One can see that the breakdown occurs at a voltage level of approximately 20 kV with a corresponding peak current of ~18 A. Note, the double peaks seen on the current waveform can be explained by output variations of the pulser. The sudden drop of resistance between the electrodes caused by the breakdown prevents creation of an ideal boxcar voltage waveform by the pulser and leads to appearance of the second current peak. The peak is associated with the discharge stage when highly conductive plasma channel connects the electrodes, leading correspondingly to high discharge current and low discharge voltages (typical for arcs and sparks) [17] [20].

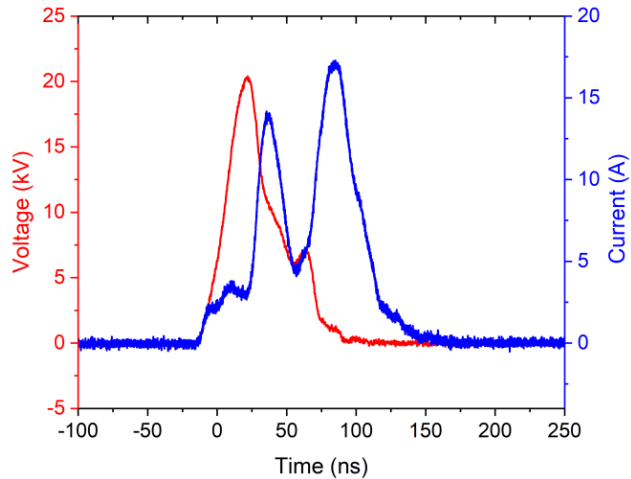

**Figure 2. Voltage (red) and current (blue) waveforms of the nanosecond HV pulse.**

The volume of the post-discharge plasma channel is approximated from ICCD imaging, and the temporal evolution of the column diameter and volume is plotted in **Figure *3***(a). One can see that the volume of the plasma expands for 0.5-9 µs [21]. A smoothed line connecting datapoints will be utilized below for the evaluation of $n_e$ from the microwave scattering signal. **Figure *3***(b) presents a sample of ICCD images used for the evaluation of the plasma diameter. Note, some recent works utilize Abel inversion in order to evaluate the actual plasma size [22] [23]. However, spatial distribution of the photon emission might be not representative of the plasma electron spatial distribution since various sources (species) contribute to the total emission pattern. Additionally, plasma size estimated based on line-of-sight integrated images and Abel inverted ones are comparable [23]. Therefore, in this work we pursued just a rough estimation of the plasma volume obtained directly from the ICCD images without Abel inversion.

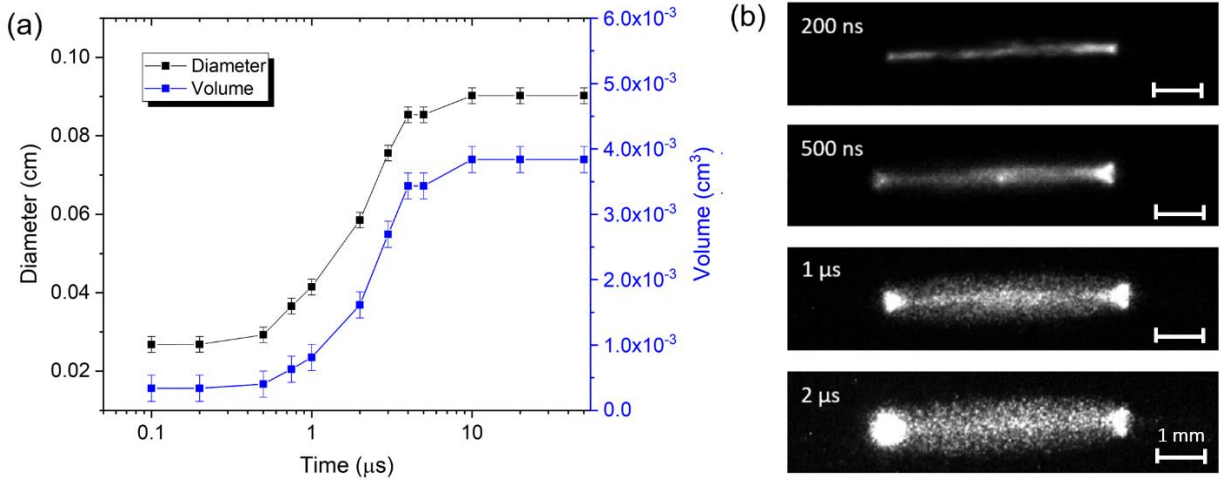

**Figure 3. (a) Temporal evolution of the plasma diameter and volume. Black line + scatter: visual measurement for diameter; Blue line + scatter: calculated plasma volume. (b) Sample ICCD images of the plasma dimensions.**

The temporal evolution of the microwave scattering signal $U_{out}$ is depicted below in **Figure *4***. It can be observed that the curve is composed of three distinct regions: 1.) fast signal decay for a duration of ~ 50 ns; 2.) a 'plateau' region for a duration of ~ 500 ns; 3.) slow signal decay for a duration of ~ 4 µs. This may appear unphysical if the decay curve for $n_e$ shares a similar pattern to $U_{out}$.

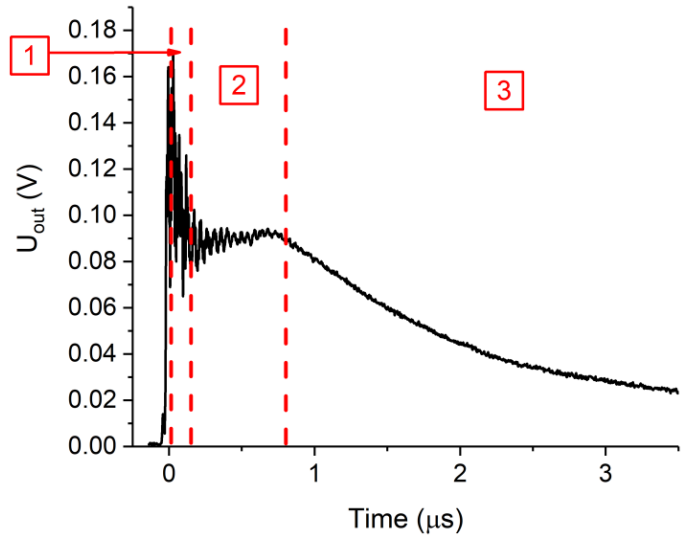

**Figure 4. Temporal evolution of $U_{out}$ measured by the MRS system. Three distinct regions of microwave scattering behavior are distinguished by the dashed red lines.**

To qualitatively understand the decay behavior of $U_{out}$, dynamics of gas density must be considered. The collisional frequency in Equation (1) can be determined as follows [20]:

$$\nu[s^{-1}] = n_g \sigma v_{Te} = 1.95 \times 10^{-10} \times n_g[cm^{-3}] \times \sqrt{T_e[K]} \qquad (2)$$

Thus, the scaling between $U_{out}$, $n_g$, $n_e$, and $V$ can be acquired using Equations (1) and (2):

$$U_{out} \propto \frac{n_e V}{n_g} \propto \frac{N_e}{n_g} \qquad (3)$$

In cases where collisional frequency cannot be assumed constant and $T_e$ is invariant, $U_{out}$ is a direct measurement of the ratio between the total electron number and local gas density. If the volume is additionally constant throughout the decay process, $U_{out}$ is then the measurement of the ratio of electron number density over local bulk gas density, which is the ionization degree (for weakly ionized plasmas).

Let us start with a qualitative interpretation of the temporal evolution of $U_{out}$ observed in **Figure 4**. For the discharge event shown in **Figure 4**, no significant changes in gas density and plasma volume are expected for the fast timescale of region 1 (t < 50 ns) – thus $n_e$ is governing change of $U_{out}$ in expression (3) and the two share a similar profile. For t > 50 ns, due to the energy deposition by the discharge, air is heated at the core and forms a shock/sonic wave that expands outwards - resulting in a local density drop. The plateau of $U_{out}$ observed for 50 ns < t < 700 ns (region 2) can thus be explained by a decrease in gas density and increase in plasma volume. Indeed, as shown in expression (3), a drop in $n_g$ and increase in $V$ can compensate for the drop in $n_e$ - resulting in the plateau. The slow decay rate observed within region 3 can also be explained by a decrease in local gas density, which will be later discussed in detail. A similar dynamic change in local gas density post-discharge can be observed in others' work: namely, in a ns-discharge with pin-to-plane geometry at atmospheric condition, simulations show that local gas density drops up to 2-times 700 ns after the discharge, which is consistent with the time period of region 2 observed. After reaching the minimum value, the gas density slowly recovers and remains approximately constant for the next few microseconds [21].

Let us now conduct quantitative studies of the local gas number density and determine $N_e$ and $n_e$. The local relative gas density ($n_g/n_{g0}$, where $n_{g0}$ is gas number density at standard atmospheric conditions -2.5×10$^{19}$ cm$^{-3}$) was acquired by Laser Rayleigh scattering from 150 ns up to 1 ms after the discharge. The resulting temporal evolution of the local relative gas density at the core of the discharge is presented in **Figure 5**(a) below. One can see that for ~ 1 μs after the discharge, local $n_g$ quickly drops to 30% of the ambient level and stays constant until about 50 μs. It then slowly recovers to the ambient level within the next ~900 μs. **Figure 5**(b) shows some example profiles of the gas density distribution at the mid-point between the electrodes along the radial direction for 1-5 μs post-discharge. A shock/sonic wave that propagates outwards at a velocity of approximately 500 m/s can also be observed. A smoothed line connecting data points for the first 5 μs was implemented for the determination of $N_e$ and $n_e$ from the MRS measurements. Note that the LRS signal is typically intricate and entangled with straylight, reflection, fluorescence, Thomson scattering, and Raman scattering. However, straylight, reflection, and fluorescence are mitigated through the aforementioned polarizer, bandpass filter, spectra absorbing backdrop, and short ICCD gate – as confirmed by negligible signals off the main illumination line and without the laser. Further, Thomson scattering is low in comparison with the pure LRS signal due to a low ionization degree ($\alpha < 10^{-3}$ and $n_e < 5\times10^{15}$ cm$^{-3}$) used in this study. Indeed, by estimating intensity of both Thomson and Rayleigh laser scattering based on known cross-sections and experimentally determined number densities, and by reviewing experimental results reported previously, it can be inferred that contribution of Thomson scattering signal is important only when $n_e \geq 10^{16}$ cm$^{-3}$ [16] [23]. Raman scattering is also distinguishably weaker than Rayleigh scattering as justified through a smaller cross section (~ 2 orders of magnitude) [16]. Such estimates are consistent with efforts to disentangle these events and have been consequently considered via an uncertainty bar. In addition, the perturbation to LRS signal caused by gas dissociation (e.g., nitrogen and oxygen) is negligible due to low degree of dissociation expected in conditions of current work based on experimental and computation data reported previously [18] [19] [22] [24].

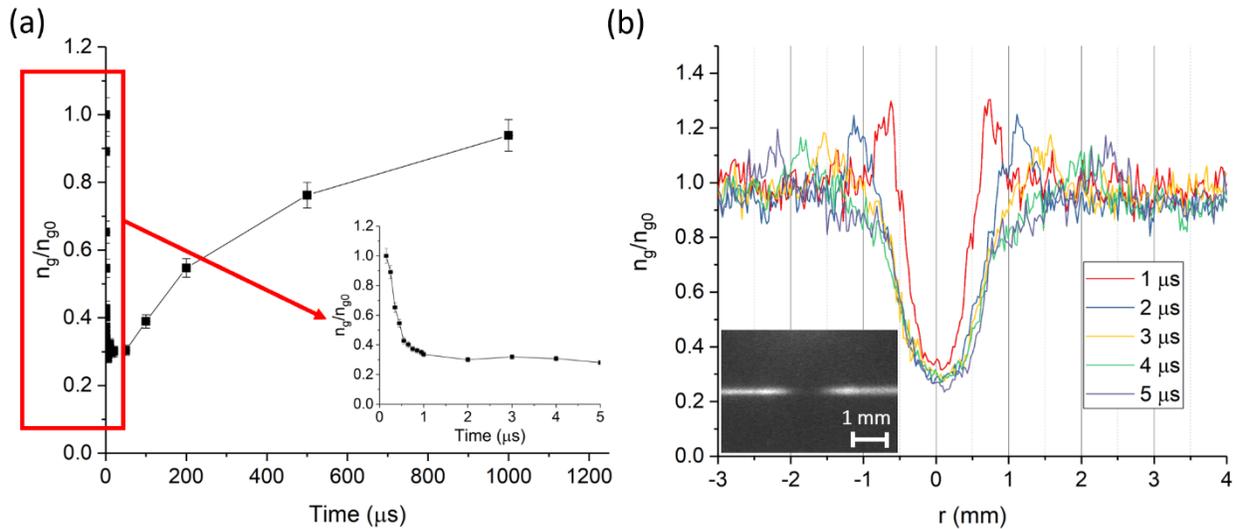

**Figure 5. (a) Temporal evolution of local gas density measured by LRS. (b) Example profiles of the density distribution along the radial direction of plasma filament for 1-5 µs after discharge. A sample of the LRS raw image is displayed on the left-bottom corner at t = 3 µs.**

With the measured temporal evolution of local gas density, one can calculate the collisional frequency $v$ with equation (2) and finally determine $N_e$ and $n_e$ from the MRS measurements. (The electron temperature $T_e$ was assumed to be decaying from 10000 K down to 2000 K within the initial 1 µs and stays constant at 2000 K [25]) The calculated $v$ is substituted into equation (1) to acquire the temporal evolution of $N_e$ and the corresponding result is shown below in **Figure** *6*(a). One can see from **Figure** *6*(a) that the maximum number of electrons produced is approximately $1.0×10^{12}$. The correction performed with LRS 'smooths' the 'plateau' observed in **Figure 4** – supporting the previous physical explanation. It additionally lowers the relative values in region 3 of **Figure 4**, reducing the decay time of $N_e$ to around 1 µs. Finally, after substituting the ICCD-measured volume of the plasma, one can acquire the corrected temporal evolution of $n_e$ which is additionally illustrated in **Figure 6**(b). One can see that the peak value of $n_e$ reaches approximately $4.3×10^{15}$ cm$^{-3}$ and decays around ~500 ns. In comparison with decay curve of $N_e$, the decay of $n_e$ is faster. This is due to the factor of plasma volume expansion which further enhances the reduction of electron number density.

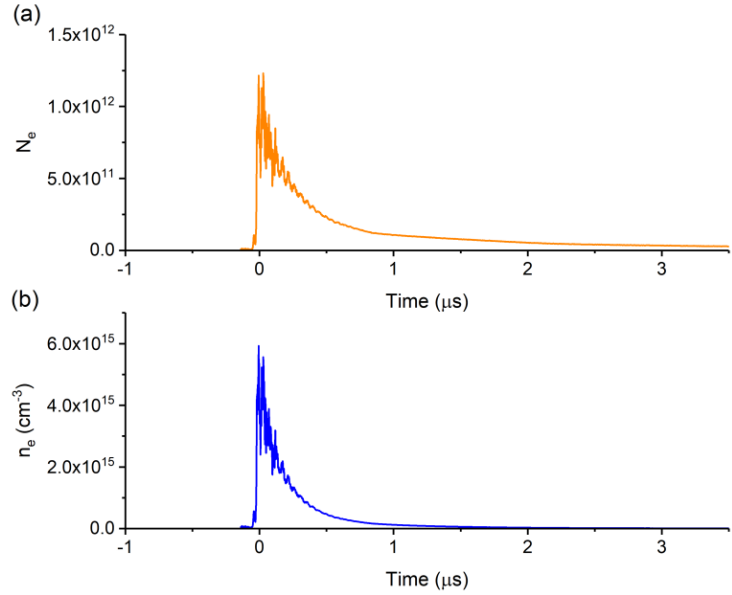

**Figure 6. Temporal evolution of (a) $N_e$ and (b) $n_e$.**

Let us now examine the experimental decay curve with theoretical predictions. Two mechanisms can contribute to the decay of $n_e$: ion-electron recombination ($\beta$ as the recombination rate) and three-body attachment ($v_{att}$ as the attachment rate). One can see that the peak value of $n_e$ was found to be $n_e \sim 4.3×10^{15}$ cm$^{-3}$. Since three-body attachment is expected to be dominant when $n_e < 10^{16}$ cm$^{-3}$ (where $\beta·n_e < v_{att}$), the governing decay mechanism in our case can be determined to be three-body attachment: $n_e(t) = n_{e_0} e^{-v_{att} t}$, where $v_{att.}$ can be expressed as $v_{att} = k_1 n_{N_2} n_{O_2} + k_2 n_{O_2}^2$ [26] [27] [28]. With attachment coefficient $k_1$ and $k_2$ being inversely proportional to $T_e$, one can calculate the attachment rate by:

$$v_{att}[s^{-1}] = 0.78 \times 10^8 \times \left(\frac{n_g[cm^{-3}]}{2.5·10^{19}}\right)^2 \times \left(\frac{300}{T_e[K]}\right) \qquad (4)$$

The resulting attachment rate $v_{att}$ was calculated to be in the range from $0.35×10^7$ s$^{-1}$ down to $0.1×10^7$ s$^{-1}$ as decay progresses. The corresponding characteristic decay time is around 100s of ns which is consistent with the experimental results. Further, in comparison with the value at standard atmospheric conditions ($0.78\times10^8$ s$^{-1}$), the decay rate in our case is lower by 20-70 times due to local reduction of neutral gas density at the discharge location.

To conclude, in our recent studies of pin-to-pin ns-discharges utilizing Rayleigh microwave scattering diagnostics, we encountered cases where the MRS signal cannot be directly interpreted into electron number density. This is attributed to a dynamic change in local air properties – such as gas density and electron temperature. Through the application of laser Rayleigh scattering, local gas density was measured and utilized to 're-calibrate' the MRS system. The electron number density was then re-evaluated. For the ns-discharge plasma studied in this case, it is shown that energy deposition by the discharge reduces local gas density by as much as 30% of ambient conditions. This leads to a reduction of the three-body attachment rate by as much 70 times in magnitude. It can then be concluded that there is a necessity for local gas density and electron temperature measurements in order to fully calibrate MRS electron number/density measurements for high energy deposition discharges. A hybrid MRS and LRS diagnostic technique demonstrated in this work can be utilized for a broad spectrum of atmospheric-pressure microplasmas where a variation in number gas density is expected due to an energy deposition in the discharge such as atmospheric pressure plasma jets and laser-induced plasmas  [29] [30] [31].

**Acknowledgements**
We thank N. A. Popov, M. N. Shneider, S. Bane, B. Singh, and L. Rajendran for useful discussions. This work was supported by the U.S. Department of Energy (grant no. DE-SC0018156) and partially by the DOE National Science Foundation (Grant No. 1903415).

# References

[1] A. Shashurin, M. N. Shneider, A. Dogariu, R. B. Miles and M. Keidar, "Temporary-resolved measurement of electron density in small atmospheric plasma," *Appl. Phys. Lett.,* vol. 96, no. 171502, pp. 1-3, 2010.

[2] X. Wang and A. Shashurin, "Study of atmospheric pressure plasma jet parameters generated by DC voltage driven cold plasma source," *J. Appl. Phys.,* vol. 122, no. 063301, 2017.

[3] A. Sharma, M. N. Slipchenko, M. N. Shneider, X. Wang, K. A. Rahman and A. Shashurin, "Counting the electrons in a multiphoton ionization by elastic scattering of microwaves," *Sci. Reports,* vol. 8, no. 2874, 2018.

[4] M. N. Shneider and R. B. Miles, "Microwave diagnostics of small plasma objects," *J. Appl. Phys.,* vol. 98, no. 033301, 2005.

[5] X. Wang, P. Stockett, R. Jagannath, S. Bane and A. Shashurin, "Time-resolved measurements of electron density in nanosecond pulsed plasmas using microwave scattering," *Plasma Sources Sci. Technol.,* vol. 27, no. 07LT02, 2018.

[6] X. Wang and A. Shashurin, "Study of the transition between modes of nanosecond repetitive pulsed discharge," in *AIAA Aviation*, Dallas, TX, 2019.

[7] D. V. Roupassov, A. A. Nikipelov, M. M. Nudnova and A. Y. Starikovskii, "Flow seperation control by plasma actuator with nanosecond pulsed-periodic discharge," *AIAA Journal,* vol. 47, p. 1680185, 2009.

[8] J. Little, K. Takashima, M. Nishihara, I. Adamovich and M. Samimy, "Separation control with nanosecond-pulse-driven dielectric barrier discharge plasma actuators," *AIAA Journal,* vol. 50, pp. 350-365, 2012.

[9] M. S. Bak, W. Kim and M. A. Cappelli, "On the quenching of excited electronic states of molecular nitrogen in nanosecond pulsed dishcarges in atmospheric pressure air," *Appl Phys Lett,* vol. 98, 011.

[10] W. Kim, M. G. Mungal and M. A. Capelli, "The role of in situ reforming in plasma enhanced ultra lean premixed methan/air flames," *Combustion and Flame,* vol. 157, pp. 374-383, 2010.

[11] X. Wang and A. Shashurin, "Gas thermometry by optical emission spectroscopy enhanced with probing nanosecond plasma pulse," *AIAA Journal,* vol. 58, no. 7, 2020.

[12] R. B. Miles, W. R. Lempert and J. N. Forkey, "Laser Rayleigh scattering," *Meas. Sci. Technol.,* vol. 12, pp. R33-R51, 2001.

[13] M. Boyda, G. Byun and K. T. Lowe, "Investigation of velocity and temperature measurement sensitivities in cross-correlation filtered Rayleigh scattering (CCFRS)," *Meas. Sci. Technol.,* vol. 30, no. 044004, 2019.

[14] R. L. McKenzie, "Progress in Laser Spectroscopic Techniques for Aerodynamic Measurements: An Overview," *AIAA Journal,* vol. 31, no. 3, 1993.

[15] F.-q. Zhao and H. Hiroyasu, "The application of laser Rayleigh scattering to combustion diagnostics," *Prog. Energy Combst. Sci.,* vol. 19, pp. 447-485, 1993.

[16] A. F. H. v. Gessel, E. A. D. Carbone, P. J. Bruggeman and J. J. A. M. v. d. Mullen, "Laser scattering on an atmospheric pressure plasma jet: disentagling Rayleigh, Raman and Thomson scattering," *Plasma Sources Sci. Technol.,* vol. 21, no. 015003, 2012.

[17] X. Wang, R. Jagannath, S. Bane and A. Shashurin, "Experimental study of modes of operation of nanosecond repetitvely pulsed discharges in air," in *AIAA Scitech*, San Diego, 2019.

[18] D. Pai, G. Stancu, D. Lacoste and C. Laux, "Nanosecond repetitively pulsed discharges in air at atmospheric pressure - the glow regime," *Plasma Sources Sci. and Technol.,* vol. 18, no. 045030, 2009.

[19] C. Limbach, C. Dumitrache and A. P. Yalin, "Laser light scattering from equilibrium, high temperature gases: limitations on Rayleigh scattering thermometry," in *AIAA Plasmadynamics and lasers conference*, Washington, D.C., 2016.

[20] Y. P. Raizer, Gas Discharge Physics, Berlin: Springer-Verlag, 1991, pp. 350 - 352.

[21] D. A. Xu, M. N. Shneider, D. A. Lacoste and C. O. Laux, "Thermal and hydrodynamic effects of nanosecond discharges in atmosphreic pressure air," *J. Phys. D: Appl. Phys,* vol. 47, no. 235202, 2014.

[22] D. L. Rusterholtz, D. A. Lacoste, G. D. Stancu, D. Z. Pai and C. O. Laux, "Ultrafast heating and oxygen disssociation in atmospheric pressure air by nanosecond repetitively pulsed discharges," *J. Appl. D: Appl. Phys,* vol. 46, no. 464010, 2013.

[23] C. M. Limbach, *Characterization of nanosecond, femtosecond and dual pulse laser energy deposition in air for flow control and diagnostic applications,* PhD Thesis, 2015.

[24] E. I. Mintoussov, S. J. Pendleton, F. G. Gerbault, N. A. Popov and S. M. Starikovskaia, "Fast gas heating in a nitrogen-oxygen discharge plasma: II. Energy exchange in the afterglow of a volume nanosecond discharge at moderate pressure," *J. Phys. D: Appl. Phys.,* vol. 44, no. 285202, 2011.

[25] J. Miles, C. Murray, A. Ross, K. Lemmer, J. Russell and S. Adams, "Time resolved electron density and temperature measurements via Thomson scattering in an atmospheric nanosecond pulsed discharge," *Plasma Sources Sci. Technol.,* vol. 29, no. 07LT02, 2020.

[26] A. Dogariu, M. N. Shneider and R. B. Miles, "Versatile radar measurement of the electron loss rate in air," *Appl. Phys. Lett.,* vol. 103, no. 224102, 2013.

[27] J. J. Jennon and M. J. Mulcapy, "Microwave measurement of attachment in oxygen-nitrogen mixtures," *Proc. Phys. Soc.,* p. 78, 1543.

[28] D. Spence and G. J. Schulz, "Three-body attachment in O2 using electron beams," *Physical Review A,* vol. 5, no. 2, 1972.

[29] X. Deng, A. Y. Nikiforov, P. Vanraes and C. Leys, "Direct current plasma jet at atmosphreic pressure operating in nitrogen and air," *J. Appl. Phys.,* vol. 113, no. 023305, 2013.

[30] M. Thiyagarajan and J. Scharer, "Experimental investigation of ultraviolet laser induced plasma density and temperature evolution in air," *J. Appl. Phys.,* vol. 104, no. 013303, 2008.

[31] A. Sharma, E. L. Braun, A. R. Patel, K. A. Rahman, M. N. Slipchenko, M. N. Shneider and A. Shashurin, "Diagnostics of CO concentration in gaseous mixtures at elevated pressures by resonance enhanced multi-photon ionization and microwave scattering," *J. Appl. Phys.,* vol. 128, no. 141301, 2020.